\documentclass[3p,times]{elsarticle}

\usepackage{xcolor}
\usepackage{amsmath}
\usepackage{amsfonts}
\usepackage{floatrow}
\usepackage{graphicx}
\usepackage[label font=normalfont,labelformat=parens]{subfig}
\usepackage{caption}
\newcommand{\beqn}{\begin{equation}}
\newcommand{\eeqn}{\end{equation}}

\newcommand{\bA}{{\bf A}}

\newcommand{\bI}{{\bf I}}

\newcommand\bLambda{\boldsymbol \Lambda}

\newcommand\bq{\boldsymbol q}

\newcommand\bu{\boldsymbol u}
\newcommand\bU{\boldsymbol U}
\newcommand\bv{\boldsymbol v}
\newcommand\bV{\boldsymbol V}

\newcommand\bW{\boldsymbol W}

\newcommand\bo{\boldsymbol 0}

\newcommand\bS{\boldsymbol S}
\newcommand\bT{\boldsymbol T}

\newcommand\bSigma{\boldsymbol{\Sigma}}

\newcommand\bR{\boldsymbol{R}}

\usepackage{amsmath}
\usepackage{amssymb}
\usepackage{float}
\biboptions{sort&compress}

\usepackage{multirow}
\title{Higher order dynamic mode decomposition to model reacting flows}
\author[UPM]{Adrián Corrochano\corref{cor1}}
\author[ULB,POLIMI]{Giuseppe D'Alessio}
\author[ULB]{Alessandro Parente}
\author[UPM]{Soledad Le Clainche}

\cortext[cor1]{Corresponding author}

\address[UPM]{School of Aerospace Engineering, Universidad Politécnica de Madrid, 28040 Madrid, Spain}
\address[ULB]{Universit\'{e} Libre de Bruxelles, \'{E}cole polytechnique de Bruxelles, Aero-Thermo-Mechanics Laboratory, Bruxelles, Belgium }
\address[POLIMI]{CRECK Modeling Lab, Department of Chemistry, Materials and Chemical Engineering, Politecnico di Milano, Piazza Leonardo da Vinci 32, 20133 Milano, Italy }

\begin{document}
	
\begin{abstract}
	In this work, the application of the multi-dimensional higher order dynamic mode decomposition (HODMD) is proposed for the first time to analyse combustion databases. In particular, HODMD has been adapted and combined with other pre-processing techniques (generally used in machine learning), in light of the multivariate nature of the data. A truncation step separate the main dynamics driving the flow from less relevant non-linear dynamics. The method is applied to analyse a database obtained from a Computational Fluid Dynamics (CFD) simulation of an axisymmetric, time varying, non-premixed, co-flow methane flame carried out by means of a detailed kinetic mechanism. Results show that HODMD can reconstruct the main jet dynamics with a reduced number of relevant modes, able to reproduce the system dynamics. These modes are found to be representative for the main flow physics with two main advantages: (i) they provide for the possibility to achieve a strong simplification with respect to the high-dimensional input data, and at the same time (ii) a small reconstruction error with respect to the original dataset is observed. In addition, the method was also validated considering a reduced matrix obtained using Principal Component Analysis (PCA) based feature selection and the Varimax rotation. This validation also reveals that it is not important to have all the variables in the dataset, just a group of them is necessary to obtain the main dynamics of the system. This has an impact on feature selection and on the cost these methodologies for very massive data.
		
\end{abstract}


\maketitle

\section{Introduction}
\label{sec:intro}

Fighting the climate change is one of the major challenges of our society. Reducing the pollutants produced from the combustion of fossil fuels, improving the energy efficiency in industrial devices, or producing new smart energy carriers, are only few examples from the multiple research topics in this field. A strong effort is carried out by the research community to study in detail and to understand the flow physics of reacting flows, motivated by the need to improve the performance and the efficiency of the combustion process and with the aim at finding new eco-friendly alternatives to the current use of fossil fuels.
 
However, the high complexity associated to the combustion process, from both a physical and chemical point of view, makes it challenging to perform detailed experimental or numerical investigations. 
On the one hand, carrying out experimental measurements is very complicated because of the small characteristic time scales and the complex chemical reactions driving the dynamics in reacting flows, and also because of the high cost to the use of specialized and sophisticated infrastructures.
On the other hand, the large number of species involved in combustion processes, the range of scales and the non-linear turbulence-chemistry impose the use of large computer facilities to obtain accurate and extensive databases allowing to study in detail the complex physics of the flow to understand turbulent combustion \cite{cant2002high}. Moreover, the enormous computational cost (CPU time and memory) associated to solve industrial problems limits the advancement in the field. A good alternative to overcome such limitations is using reduced order models (ROMs). ROMs have lately spread within the field of reacting flows, as they offer the opportunity to simplify the complex nature of the physical problem, as well as to create high fidelity lower-dimensional models which represent the physics with a low level of error \cite{parente2011investigation,bellemans2018feature,Coussement2013}. Using ROMs, it is possible to model the main flow dynamics with a reduced computational cost.

Principal Component Analysis (PCA) \cite{PCA, isaac2014reduced, parente2009identification, parente2011investigation} has found widespread application in combustion. PCA is a statistical technique whose aim is to find lower-dimensional manifolds by identifying correlations between the input variables, which are then combined linearly to eventually reduce the problem dimension \cite{bishop2006pattern, jolliffe2016principal}. 
The application of PCA has been proved suitable for dimensionality reduction and feature extraction \cite{parente2009identification, parente2011investigation, d2021feature, bellemans2018feature}, clustering \cite{d2020adaptive, d2020impact}, and data analysis and feature selection \cite{d2020analysis, jolliffe1972discarding, jolliffe1973discarding, krzanowski1987selection, jolliffe2002choosing}.
\par

More recently, dynamic mode decomposition (DMD) \cite{Schmid10} has also been introduced as a good tool to extract patterns and develop ROMs in reacting flows. The main advantage of this technique lies in its ability to identify the main frequencies and features, known as DMD modes, leading the flow dynamics of the problem studied. Such modes could be related to flow instabilities reflecting the underlying physical mechanisms driving the flow \cite{LeClaincheSOCO19}. Hence, DMD can potentially complement PCA, while considering physical principles governing the flow to develop new efficient ROMs. Some examples of DMD applied to the analysis of reacting flows in the last decade are detailed as follows. Richecoeur {\em et al.} \cite{Richecoeur2012} applied DMD to study the main dynamics in turbulent combustion, studying in detail the performance of DMD when applied to the analysis of experimental databases. Souvick {\em et al.} \cite{Souvick2013} used DMD in experimental databases to identify the main patterns connected to flow stabilities in a swirl-stabilised dump combustor. In the same line, several authors have used DMD to identify the main flow patterns, linked with the main instabilities present in the flow. For instance, in the analysis of experimental databases, Quinlan {\em et al.} \cite{Quinlan2014} developed a technique based on DMD to extract the main patterns and frequencies related to the resonant modes producing an induced transverse combustion instability; Huang {\em et al.} \cite{Huang2016}  used DMD to study of self-excited longitudinal combustion instabilities in laboratory-scaled single-element gas turbine and rocket combustors. Motheau {\em et al.} \cite{Motheau2014} used DMD for the analysis of a LES database, with the aim at developing a low-order model showing that one of the mechanism triggering combustion instability is related to convected and acoustic entropy waves. More applications of DMD to identify patterns and flow instabilities in reactive flows include the work by Abou-Taouk {\em et al.} \cite{AbouTaouk2015}, who studied a V-flame in an afterburner-type configuration; the work by Ghani {\em et al.} \cite{Ghani2015}, who identified the main patterns in an industrial gas turbine combustion chamber; and the work by Grenga {\em et al.} \cite{grenga2018dynamic, grenga2020dynamic}, who proposed a more efficient implementation of DMD, suitable to elucidate multiphysics requiring with a reduced memory usage, which was tested in a DNS database solving a turbulent premixed flame problem.

As explained in Ref. \cite{LeClaincheVegaComplexity18}, the application of DMD is limited to some specific conditions of the database analysed and the physical mechanism underlying the problem under study. Hence, in the analysis of complex dynamical systems, noisy experimental databases, DMD suffers in identifying robust and accurate results modelling the flow dynamics, or may even fail. Higher Order Dynamic Mode Decomposition (HODMD) \cite{LeClaincheVega17} is introduced as an alternative to overcome such limitations. HODMD is a more robust extension of DMD, which has been proven suitable to the analysis of non-reactive complex flows and noisy experimental data in several complex and industrial applications \cite{Corrochano,LeClaincheetalAIAA17,LeClaincheetalJAircraft18,LeClaincheetalJFM2020}. The performance of HODMD showed in the previous applications, makes it a suitable tool for the analysis of reactive flows and turbulent combustion. This article introduces for the first time, to the authors knowledge, the application of HODMD to analyse a combustion database. More specifically, HODMD algorithm is extended for the analysis of reacting flows, formed by a large number of variables and species, with different magnitudes. The method is able to reduce the data dimensionality and to identify the main frequencies and associated patterns driving the flow dynamics, with the aim at developing a ROM. This application of HODMD is illustrated in a numerical database of a laminar reacting flow representative for the oxidation of a methane jet in air. The results are validated with reduced dimensionality data obtained with PCA. Since HODMD is fully data-driven, this algorithm could be easily extended to the analysis of other type of flows.

The article is organised as follows.  Section \ref{sec:Meth} introduces the methodology used to develop this work. The description of the database generated numerically is introduced in Section \ref{sec:database}. Finally, the main results and conclusions are presented in Section \ref{sec:Results} and Section \ref{sec:conclusions}, respectively.

\section{Methodology \label{sec:Meth}}

This section introduces the theoretical aspects of the two algorithms that are applied in this paper to identify the main patterns and frequencies driving the flow dynamics: the higher order dynamic mode decomposition \cite{LeClaincheVega17} and the principal component analysis (PCA) \cite{PCA}, both combined with pre-processing strategies generally used in machine learning. In particular, PCA will also be coupled with the Varimax rotation method \cite{kaiser1958varimax} to identify a subset of main variables from a database containing a larger number of variables.

\subsection{Higher Order Dynamic Mode Decomposition \label{sec:HODMD}}
Higher order dynamic mode decomposition (HODMD) \cite{LeClaincheVega17} is a data-driven method introduced as a robust extension of dynamic mode decomposition (DMD) \citep{Schmid10} for the analysis of complex non-linear dynamical systems.
Similarly to DMD, HODMD decomposes spatio-temporal data $\bv(x,y,z,t_{k})$, collected at time instant $t_k$ (for convenience expressed as $\bv_{k}$), as an expansion of $M$ Fourier-like modes. For two-dimensional databases (as presented in this article), the decomposition reads
    \begin{equation}
 \bv(x,y,t_{k})\simeq  \bv_k^\text{approx}=
 \sum_{m=1}^M a_{m}\bu_m(x,y)e^{(\delta_m+i \omega_m)t_k},\quad
	k=1,\ldots,K,\label{ab00}
  \end{equation}
 where $\bu_m$ are the DMD modes and $a_{m}$, $\omega_m$, $\delta_m$ are their associated amplitudes, frequencies and growth rates. From a practical point of view, it is necessary to collect a group of data equi-distant in time with time interval $\Delta t$, in the following snapshot matrix
  \beqn
  \bV_1^K = [\bv_{1},\bv_{2},\ldots,\bv_{k},\bv_{k+1},\ldots,\bv_{K-1},\bv_{K}].\label{ab0}
  \eeqn
with dimension $J\times K$, with $J=N_{variables}\times N_x \times N_y$, where $N_{variables}$ are the number of variables considered in the database, and $N_x$ and $N_y$  are the number of grid points along the streamwise and normal (or radial) directions. Hence, for a vector containing $j$ variables, each variable, formed by  vectors collected at time $t_k$, is included in consecutive rows in the snapshot matrix (\ref{ab0}). In what follows, when the snapshot matrix $\bV_1^K$ formed by a single variable $j$, it is referred as $\bv_j$.

HODMD relies on the following high-order Koopman assumption, which relates $d$ (tunable) subsequent snapshots as
\begin{equation}
	\bv_{k+d} = \textbf{R}_1\bv_{k} +   \textbf{R}_2\bv_{k+1} + \ldots  \textbf{R}_d\bv_{k+d-1}  ,\quad
	k=1,\ldots,K-d,\label{e23}
\end{equation}
where $\textbf{R}_i$ (for $i=1,...,d$) are the (so-called)  Koopman operators, which are linear and contain the dynamics of the system. This is the essence of HODMD, which provides high accurate results, even in the analysis of highly complex databases. When $d=1$, the Koopman assumption approximates the standard DMD algorithm. Details about the HODMD algorithm can be found in \cite{LeClaincheVega17,LeClaincheVegaSoria17}, and the Matlab codes can be found in \cite{HODMDbook}.

In what follows, the accuracy of the approximation of the original database ($\bv_k^\text{approx}$) using the DMD expansion eq. (\ref{ab00}) will be measured in terms of the relative root
mean square (RRMS) error as
  \beqn
  \text{RRMS error}=\sqrt{\frac{\sum_ {k=1}^K\|\bv_k^\text{approx}-\bv_k\|^2_2}{\sum_ {k=1}^K\|\bv_k\|^2_2}}.\label{error}
  \eeqn

\subsection{HODMD in combustion flows \label{sec:HODMDcombustion}}
The robustness and accuracy of HODMD has been tested in a wide range of applications, including the analysis of complex flows (i.e., noisy experiments\citep{LeClaincheVegaSoria17}, transitional flows\cite{LeClaincheFerrer2018}, turbulent flows\cite{LeClaincheetalJFM2020}) to create reduced order models\citep{LeClaincheVegaPoF17} or to extract patterns and to study the flow physics\citep{LeClaincheVegaComplexity18}. However, the analysis was limited to database including the three velocity components and pressure. In combustion systems, it is common to deal with a larger number of variables with different magnitudes. Hence, it is necessary carefully pre-processing the data before starting with the analysis, with the aim at extracting the maximum quantity of relevant information from the problem studied. For such aim, this article introduces a three-steps method based on HODMD, which summarizes the novel application of this algorithm for the analysis of combustion data. Moreover, different strategies to reduce the data dimensionality are introduced and discussed. The three-step method is defined as follows.
\begin{itemize}
\item[{\bf Step 1:}] {\bf Centering and scaling.}

Combustion data usually consist of variables that have different units and ranges, such as temperature and chemical species, which therefore entails the need to pre-process them  before applying any data-driven algorithm. 
{\it Centering} consists on the subtraction of the mean value for a given variable to focus on the fluctuations.
{\it Scaling} the variables to a uniform range is also important because it is common to have order of magnitudes of difference among the different variables. The aforementioned pre-processing operations are accomplished as follows: 
\begin{equation}
	\tilde{\bv}_j = \dfrac{\bv_j-\bar{\bv}_j}{c_j}\label{e24},
\end{equation}
\noindent where $\bar{\bv}_j$ represents the temporal mean of the variable $\bv_j$, and $c_j$ is the scaling factor. In this paper, two scaling methods are considered:

\begin{itemize}
	\item Auto scaling: Uses the standard deviation of each variable, $\sigma_{j}$, as the scaling factor.
	\item Range scaling: Uses the difference between the maximum and minimum value of each variable as the scaling factor.
\end{itemize}

As observed by Parente and Sutherland \cite{PCA}, some scaling methods may emphasize the relevance of some variables because of their distribution. 
Auto scaling gives similar importance to all the different variables, while range scaling performs better for most of the major species. 

\item[{\bf Step 2:}] {\bf Dimension reduction via HOSVD.}

A two-dimensional database can be re-organized into a fourth-order tensor $\bA_{i_1i_2i_3k}$ (with indexes associated to each one of the components, for $i_1=1,\cdots, N_{variables}$, $i_2=1,\cdots, N_x$, $i_3=1,\cdots, N_y$ and $k=1,\cdots, K$ ), in which each one of the four dimensions correspond to the variables of the dataset, the two spatial components (namely, streamwise, $x$, and normal or radial, $y$, components) and the temporal component, $t$, respectively. The high order singular value decomposition (HOSVD) method \cite{HOSVD}  can then be used to reduce the data dimensionality in a highly efficient way. This algorithm applies standard SVD to each one of the fibers composing the tensor (see details and the Matlab code to perform this analysis in Ref. \cite{HODMDbook}), resulting in a expansion as
\begin{equation}
  \bA_{i_1i_2i_3k} = \sum_{p_1=1}^{P_1}\sum_{p_2=1}^{P_2}\sum_{p_3=1}^{P_3}\sum_{n=1}^{N} \bS_{p_1p_2p_3n} \bU_{i_1p_1}\bW^{(x)}_{i_2p_2} \bW^{(y)}_{i_3p_3} \bT_{kn}\label{e25}
\end{equation}

where $\bS_{p_1p_2p_3n}$ is a fourth-order tensor (called the {\em  core tensor}) and the columns of the matrices $\bU$,
$\bW^{(x)}$, $\bW^{(y)}$, and $\bT$ are known as the {\em SVD modes} of the decomposition (related to the number of components, called as the {\it component modes}, and the streamwise and normal/radial spatial dimension, called as the {\it spatial modes}, respectively).  Matrix $\bT$ corresponds to the {\it SVD temporal modes} matrix, the {\it reduced snapshot matrix} on which the HODMD algorithm is applied, as described in the next Step. 
The number of SVD modes, $N$, associated to the temporal matrix $\bT$ are calculated  based on the tolerance (tunable) $\varepsilon$  as
\begin{equation}
	 \sigma_{N+1}/\sigma_1\leq \varepsilon.\label{e26}
\end{equation}
The SVD modes, associated to the remaining matrices, $P_1$, $P_2$ and $P_3$, are calculated in a similar way for each one of the group of modes. 
Generally, in standard fluid dynamics applications, the first component of the tensor, $\bU$, is not truncated, since the number of variables is (at most) $4$ (pressure and three velocity components). However, considering the typical size of combustion applications, we also truncate this first dimension, evaluating the influence of this truncation in the reconstruction error as in eq. (\ref{error}). In this way, it is possible to identify what is the influence of the different variables in the system dynamics.

A new variable will quantify the compression carried out by the HOSVD application in the original database, the {\it compression factor} ($CF$), defined as
\begin{equation}
	CF = \dfrac{N_e}{N_{er}} \label{e41}
\end{equation}
where $N_e$ is the total number of elements of the tensor (namely, $N_e= N_{variables} \times N_x \times N_y \times K$), and $N_{er}$ is the number of elements retained after reducing the data dimensionality with HOSVD.

\item[{\bf Step 3:}] {\bf The DMD-d algorithm.}

The DMD-d algorithm is applied to the matrix associated with the temporal modes $\bT$, referred as $\widehat{\bT}_1^K$, of dimension $N \times K$, with $N$ the number of SVD temporal modes retained,  also known as {\it spatial complexity} of the system, and $K$ the number of snapshots forming the original dataset. For consistency, to compare with the algorithm described in the literature \citep{LeClaincheVegaSoria17}, this matrix is called as the  {\it reduced snapshot matrix}.
The high order Koopman assumption eq. (\ref{e23}) is applied to this snapshot matrix as
\beqn
\widehat{\bT}_{d+1}^K\simeq \widehat{\bR}_1 \widehat{\bT}_1^{K-d}+ \widehat{\bR}_2 \widehat{\bT}_2^{K-d+1} + \ldots + \widehat{\bR}_d \widehat{\bT}_d^{K-1}.\label{ab26}
\eeqn
This equation divides  the snapshot matrix into $d$ blocks, each one containing time-delayed $K-d$ snapshots.

 The previous equation is rewritten in terms of the {\it modified Koopman matrix} $\tilde{\bR}$ and the {\it modified snapshot matrix} $\tilde{\bT_1}^{K-d+1}$   as
\beqn
\tilde{\bT}_2^{K-d+1}=\tilde{\bR}\,\tilde{\bT}_1^{K-d},\label{ab30}
\eeqn
also written as
\beqn
\left[\begin{array}{c}\widehat{\bT}_2^{K-d+1}\\
\ldots\\ \widehat{\bT}_{d}^{K-1}\\\widehat{\bT}_{d+1}^{K} \end{array}\right] =
\left[\begin{array}{cccccc}
\bo&\bI &\bo&\ldots &\bo&\bo\\
\bo &\bo&\bI&\ldots&\bo &\bo\\
\ldots & \ldots & \ldots&\ldots&\ldots&\ldots \\
 \bo &\bo&\bo& \ldots&\bI &\bo\\
 \widehat{\bR}_1 &\widehat{\bR}_2&\widehat{\bR}_3& \ldots&\widehat{\bR}_{d-1}&\widehat{\bR}_d \end{array}\right]
\cdot 
\left[\begin{array}{c}\widehat{\bT}_1^{K-d}\\ \widehat{\bT}_2^{K-d+1}\\
\ldots\\ \widehat{\bT}_{d}^{K-1} \end{array}\right].
 \label{ab32}
\eeqn
This new matrix, which organizes the snapshot blocks identified in the high order Koopman assumption in columns, increases the spatial complexity of the data to a value $N'>N$, increasing in this way the accuracy in the calculations of the DMD modes. This step justifies the better performance of HODMD compared to DMD and other variants, which are all based on a Koopman expansion eq. (\ref{e23}), with $d=1$.

A new dimension-reduction is carried out into this matrix, to eliminate possible redundancies,  via
truncated SVD using again the tolerance $\varepsilon$ in the new calculated singular values as
\beqn
\tilde\sigma_{N'+1}/\tilde\sigma_{1}<  \varepsilon_1,
\label{ab37}
\eeqn
where $N'>N$ (new spatial complexity) is the number of retained SVD modes. 
Hence, the modified snapshot matrix is obtained as
\beqn
\tilde{\bT}_1^{K-d+1} \simeq \tilde{\bU}\tilde{\bSigma}\tilde{\bT}^\top \simeq \tilde{\bU}\overline{\bT}_1^{K-d+1}, \label{ab35}
\eeqn
 with  $\overline{\bT}_1^{K-d+1}=\tilde{\Sigma}\tilde{\bT}^\top$, where $\tilde{\bU}^\top\tilde{\bU} = \tilde{\bT}^\top\tilde{\bT}$ are the
$N'\times N'-$unit matrices, and the diagonal of matrix $\tilde{\bSigma}$ contains
the singular values $\tilde{\sigma}_1,\cdots,\tilde{\sigma}_{N'}$. To complete this step, eq. (\ref{ab30}) is pre-multiplied by $\tilde{\bU}^\top$, which invoking (\ref{ab35}) results as
\beqn
\overline{\bT}_2^{K-d+1}=\overline{\bR}\,\overline{\bT}_1^{K-d}.\label{ab39}
\eeqn
The new $N'\times N'$-Koopman matrix is related to $\tilde{\bR}$ by
$\overline{\bR}\simeq\tilde{\bU}^\top\tilde{\bR}\tilde{\bU}$. Nevertheless, we use the pseudoinverse in eq.(\ref{ab39}) (via non-trucated SVD) to the matrix  $\overline{\bT}_1^{K-d}$, instead of computing the previous expresion, which yields
\beqn
\overline{\bT}_1^{K-d}=\overline{\bU} {\bLambda} \overline{\bV}^\top, \label{ab41}
\eeqn
where $\overline{\bU}\overline{\bU}^\top=\overline{\bU}^\top\overline{\bU}=\overline{\bV}^\top\overline{\bV}$ are
 the $N'\times N'-$unit matrices and the diagonal of $\bLambda$ contains the $N'$
singular values.
Substituting
  (\ref{ab41}) into (\ref{ab39}) and  post-multiplying by $\overline{\bV}{\bLambda}^{-1}\overline{\bU}^\top$, it is possible obtaining the following equation
\beqn
\overline{\bR}= \overline{\bT}_2^{K-d+1}\overline{\bV}{\bLambda}^{-1}\overline{\bU}^\top. \label{ab43}
\eeqn
The reduced DMD expansion for the reduced snapshots  is then calculated using $N'$ eigenvectors $\overline{\bq}_m$ and eigenvalues $\mu_m$ solving the eigenvalue problem of $\overline{\bR}$, as
      \begin{equation}
\widehat{\bv}_k\simeq
 \sum_{m=1}^M \widehat{a}_{m}\widehat{\bu}_m e^{(\delta_m+i \omega_m) t_k},\label{bbb1}
    \end{equation}
     for $k=1,\ldots,K$.
Retaining only the first $N$ components of the
 vectors  $\widehat{\bq}_m=\tilde{\bU}\overline{\bq}_m$, it is possible to approximate the reduced DMD modes $\widehat{\bu}_m$ (dimension $Nd$). The damping rates $\delta_m$ and frequencies $\omega_m$ are obtained as
\beqn
\delta_m + i \omega_m=\log(\mu_m)/\Delta t. \label{ab47}
\eeqn
The amplitudes $\hat a_m$  are calculated  as in optimized DMD \citep{Chenetal12}, solving a least square fitting problem in eq.(\ref{bbb1}). Invoking (\ref{e25}) and rescaling with unit-norm the modes $\bu_m$ and the amplitudes $a_m$, it is possible to obtain the  original DMD expansion (\ref{ab00}). Finally, using a second tolerance $\varepsilon_2$ (tunable), it is possible to calculate the number of $M$ DMD modes in eq. (\ref{ab00}), called as the
  {\it spectral complexity}, as follows
 \beqn
 a_{M+1}/a_1\leq \varepsilon_2. \label{b66}
 \eeqn

When the database analysed presents the transient state of a numerical simulation, an additional step is carried out to identify the leading modes driving the flow dynamics. This step consists on calculating the energy index $I_m$, defined as
\begin{equation}
	I_m = \sum_{k = 1}^{K} \lvert a_m e^{(\delta_m + i\omega_m)t_k}\rvert \;\|u_m\|_{F}^{2} \times \Delta t, \label{e42}
\end{equation}
where $\| \bullet \|_{F}^2 $ is the square of the Frobenius norm. This index weights the modes taking into account also the damping or growth of the modes in time, which is reflected in $\delta_m$. More details about this criterion, called as {\it HODMD with criterion}, are presented in \cite{Kou}.

\end{itemize}

\subsection{Principal Component Analysis and Variable Selection\label{sec:PCAmethod}}	
Principal Component Analysis (PCA) \cite{parente2009identification} reduces the dimensionality of an input matrix $\mathbf{X}$, consisting of ($K \times N_{x} \times N_{y}$) rows representative for the statistical observations of a system (e.g., $N_{x} \times N_{y}$ grid points for $K$ snapshots of a CFD simulation) and $N_{variables}$ columns for the associated observable variables, by projecting such matrix onto a lower-dimensional basis consisting of $q$ eigenvectors. 

The PCA algorithm is associated to the covariance matrix $\mathbf{C}$ of the input dataset $\mathbf{X}$, which is defined as follows
\begin{equation}
\mathbf{C} = \frac{1}{(K \times N_{x} \times N_{y})-1} \mathbf{X}^{T} \mathbf{X}.
\end{equation}
This covariance matrix can be approximated by solving an eigenvalue problem as
\begin{equation}
\mathbf{C} = \mathbf{A}\mathbf{L}\mathbf{A}^{T},
\end{equation}
where the matrices $\mathbf{A}$ and $\mathbf{L}$ contain the associated eigenvectors (in columns) and eigenvalues (in the diagonal), respectively. 
The columns of $\mathbf{A}$ are the Principal Components (PCs) and they represent geometrically the orthogonal coordinates of the lower-dimensional manifold, obtained as a linear combination of the original variables \cite{bishop2006pattern}. This matrix also represent the proper orthogonal decomposition (POD) modes, calculated by an SVD of the original dataset $\mathbf{X}$ (see details in \cite{LeClaincheSOCO19}).
The matrix $\mathbf{L}$ is instead a diagonal matrix, containing the eigenvalues $\lambda_{i}$ (with $i \in [1, N_{variables}$]) which are found in descending order of magnitude.
This last condition on the eigenvalues' order has important effects from a mathematical point of view, and it represents a key point for the dimensionality reduction.
The magnitude of each eigenvalue $\lambda_{i}$, in fact, represents the amount of input data variance being explained by the associated eigenvector. 
Consequently, if the value of $\lambda_{i}$ is large, it entails the associated PC having an important role in the lower-dimensional representation of the input data, as it is representative for a large amount of information.
On the other hand, if $\lambda_{i}$ is small, it means that the contribution of the associated PC to the  orthogonal space representation is negligible, and it can eventually be considered as noise, spatial redundancies (or in turbulent flows, it can be related to small flow scales \cite{LeClaincheetalJFM2020}).
In light of what stated above, it follows that the last PCs can be removed so that the $p$-dimensional input matrix can be expressed only by $q$ orthogonal components (with $q < N_{variables}$) associated to the highest eigenvalues, thus considering a truncated eigenvector matrix $\mathbf{A}_{q}$.
The optimal dimensionality for the PCA manifold can be set by assessing the amount of input data variance ($t_{q}$) being explained by the truncated $q$-dimensional basis \cite{d2021feature}, defined as the ratio between the cumulative sum of the first $q$ eigenvalues and the sum of all eigenvalues
\begin{equation}
t_{q} = \frac{\sum_{i=1}^{q} \lambda_{i}}{\sum_{j=1}^{N_{variables}} \lambda_{j}},
\end{equation}
and ensuring that the condition $t_{q} \approx 1$ is observed. 
The dimensionality reduction is finally accomplished by projecting the input data on the truncated basis 
\begin{equation}\label{eq:projection}
\mathbf{Z}_{q} = \mathbf{X} \mathbf{A}_{q}.
\end{equation}
After the projection, the original input can be then reconstructed leveraging the orthonormality of the eigenvectors matrix as
\begin{equation}
\mathbf{X}_{r} \approx \mathbf{Z}_{q} \mathbf{A}_{q}^{T}.
\end{equation}
\par
The PCA algorithm shown above can be used not only to perform dimensionality reduction via feature extraction (i.e., to find a new, reduced-order, representation of the original space with a manifold whose components are obtained as a linear combination of the original ones), but also to perform feature selection (i.e., to reduce the dimensionality by selecting a subset of the original variables).
In fact, PCA can effectively identify correlations among the variables defining the state space and can subsequently select the best reduced subset from the original group of input variables \cite{jolliffe1972discarding, jolliffe1973discarding, jolliffe2002choosing}.
In particular, according to Jolliffe \cite{jolliffe2002choosing}, it is possible to discard by means of a backward elimination algorithm (the B2 algorithm) the variables that are more aligned with the last PCs (i.e., the ones having the highest loadings on the last eigenvectors) for a higher correlation with the components whose associated eigenvalues are small.
\par
\begin{figure}[ht]
	\centering
	\includegraphics[height = 0.2\textwidth]{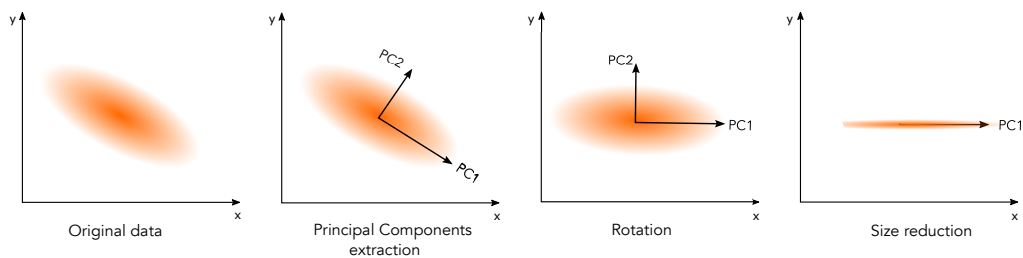}
	\caption{Schematic of the different steps for a PCA reduction process when coupled with an orthogonal rotation of the eigenvectors.}
	\label{fig:PCArot}
\end{figure}

Nevertheless, if high-dimensional problems are considered, the weights' distribution on the PCs can be difficult to analyse or interpret, as it is not always possible to identify a clearly dominant weight on one or more eigenvectors.
Thus, PCA can be coupled with a rotation technique to increase the eigenvectors' physical interpretability, for the rotated PCs to be preferentially aligned with one variable, as shown in Fig. \ref{fig:PCArot}.

Different techniques are available in literature to accomplish this operation, and they can be grouped into two classes: orthogonal and oblique rotations. 
The main difference between the latter is that when the first class of rotations is employed the eigenvectors keep their reciprocal orthogonality (and, consequently, the amount of variance being explained by the reduced basis does not change), while this condition is lost with oblique rotations.
As shown by Bellemans \emph{et al.} \cite{bellemans2018feature}, the Varimax rotation introduced by Kaiser \cite{kaiser1958varimax} can be considered the most accurate for application to reacting flows data.
In particular, this technique belongs to the class of orthogonal rotations and its objective function is the maximization of the sum of variances of the squared coefficients within each eigenvector, i.e., the maximization of the following quantity
\begin{equation} \label{eq:varimax}
\mathbf{V} = \frac{N_{variables} \sum_{i = 1}^{N_{variables}} (b_{ij}^{2})^{2} - \gamma \sum_{i = 1}^{N_{variables}} (b_{ij}^{2})^{2}}{N_{variables}^{2}},
\end{equation}
where $b$ is the PC loading, $\gamma =1$, $i \in [1, N_{variables}]$, and $j = [1, q]$ with $q$ being the dimensionality of the PCA manifold \cite{bellemans2018feature}.
When a non-unitary $\gamma$ is used, different rotation techniques are obtained (e.g., quartimax rotation when $\gamma = 0$).
\par
When Varimax rotation is applied, a new representation of the eigenvectors ($\mathbf{A}_{q}^{r}$) is obtained that can be expressed in terms of original PCs when a rotation matrix $\mathbf{T}$ is defined as in Ref. \cite{bellemans2018feature}
\begin{equation}\label{eq:rotatedPC}
\mathbf{A}_{q}^{r} = \mathbf{T} \mathbf{A}_{q},
\end{equation}
which thus entail the possibility to also retrieve a rotated lower-dimensional representation of the input data as
\begin{equation}
\mathbf{Z}_{q}^{r} = \mathbf{X} \mathbf{T} \mathbf{A}_{q}.
\end{equation}
\par
In the current work, a rotated basis as in eq. (\ref{eq:rotatedPC}) is utilized to perform feature selection via B2 algorithm (hereinafter referred to as B2r, when rotated PCs are taken into account).
The impact of the rotation on feature selection will be then compared to the case when B2 is applied to the standard PCA (that is, when an unrotated truncated basis $\mathbf{A}_{q}$ is used to discard the variables) in the following sections.

\section{Numerical database and PCA \label{sec:database}}

This section briefly introduces the numerical simulations carried out to generate the database of a reactive flow that will be analyzed using the algorithm introduced in Section \ref{sec:HODMDcombustion}. 
In particular, the considered numerical simulation is representative for the oxidation of a methane jet (65\% methane, 35\% nitrogen, on a molar basis) in air.
The oxidizer is injected into the domain adopting a constant velocity of 35 cm/s, while the fuel velocity varies in space and time according to the following sinusoidal perturbation:
\begin{equation}
v(r,t) = v_{max} \left( 1 - \frac{r^2}{R^2} \right) [1 + A \ sin(2 \pi f \ t)],
\end{equation}
with R representing the internal radius of the nozzle, r is the radial coordinate, t is the time, $v_{max}$ is the maximum velocity ($v_{max}$ =\ 70 cm/s) and f and A are the frequency (f = 10 Hz) and the amplitude (A = 0.25) of the perturbation, respectively.
With regard to the kinetic mechanism, the detailed mechanism \texttt{POLIMI\_C1\_C3\_HT\_1412} \cite{POLI} (accounting for 82 species) was used for the numerical simulations, which were run with \texttt{LaminarSMOKE} code, an \texttt{OpenFOAM}-based operator-splitting solver proposed by Cuoci \emph{et al.} \cite{cuoci2013numerical} to simulate laminar reacting flows by means of detailed kinetic mechanism.
Additional information regarding this numerical simulations and the numerical setup can be found in Refs. \cite{d2020adaptive,d2020impact}.

\par
\begin{figure}
	\centering
	\includegraphics[height = 0.4\textwidth]{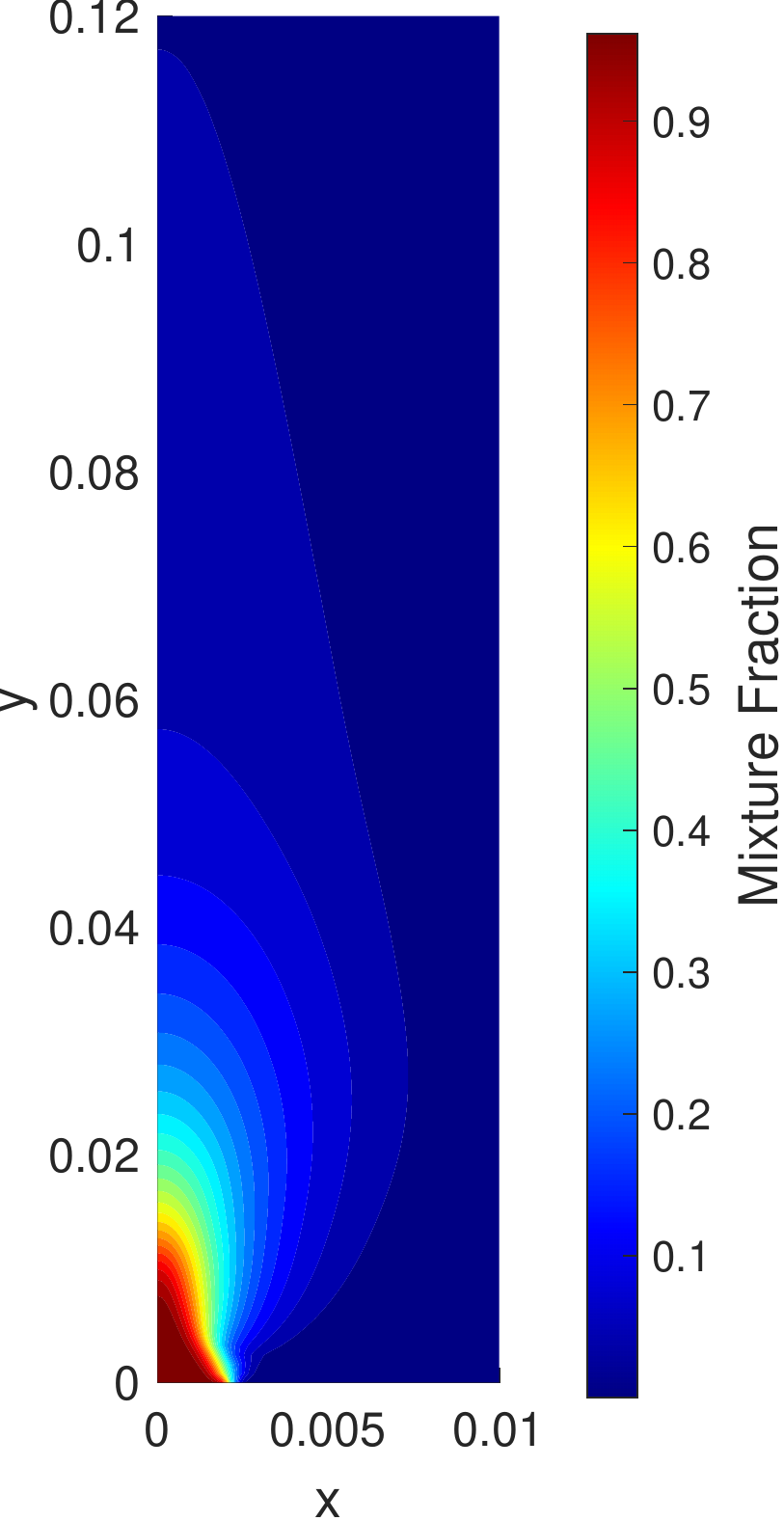}
	\hspace{0.5cm}
	\includegraphics[height= 0.4\textwidth]{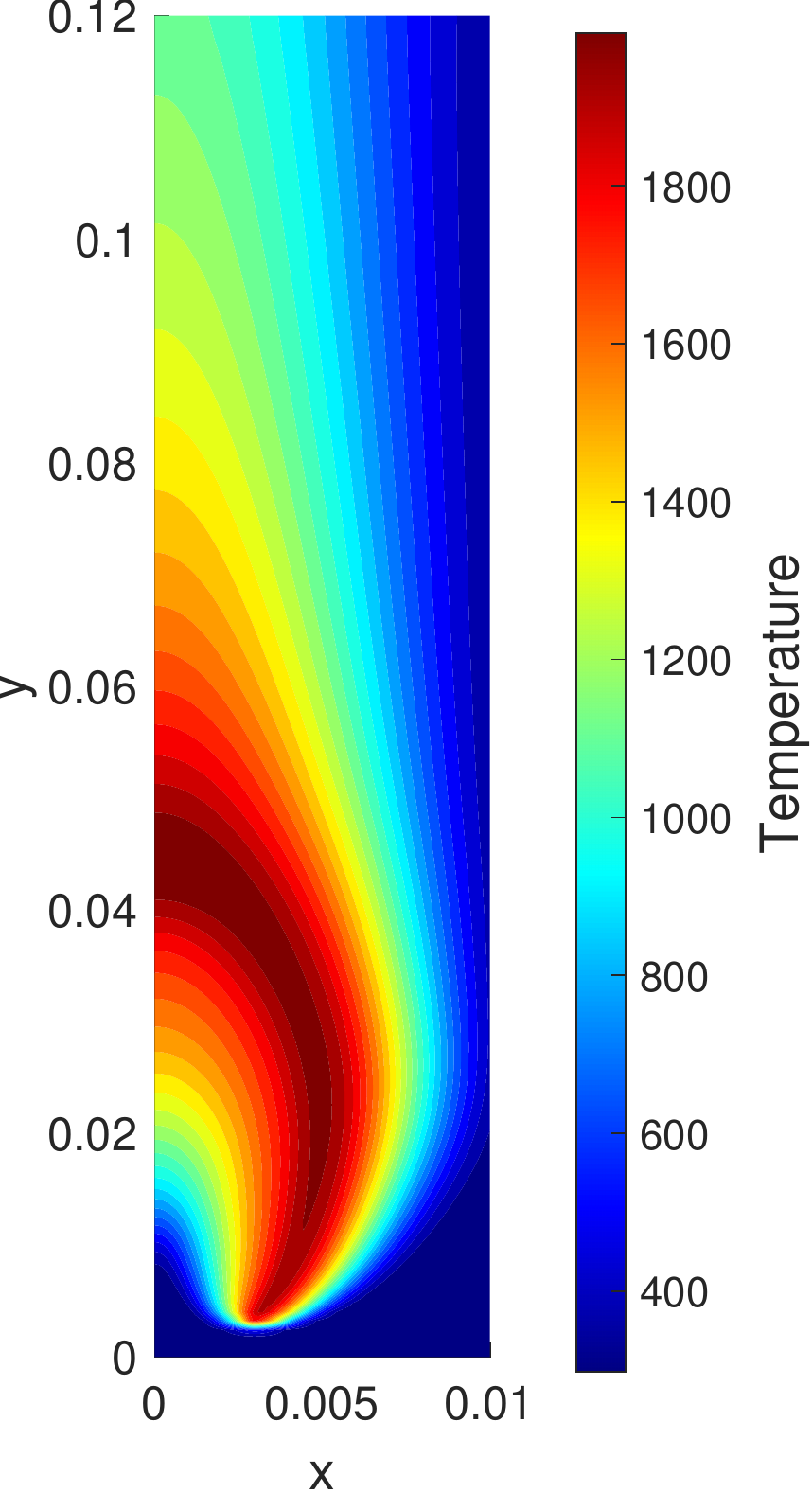}\hspace{0.5cm}
	\includegraphics[height = 0.4\textwidth]{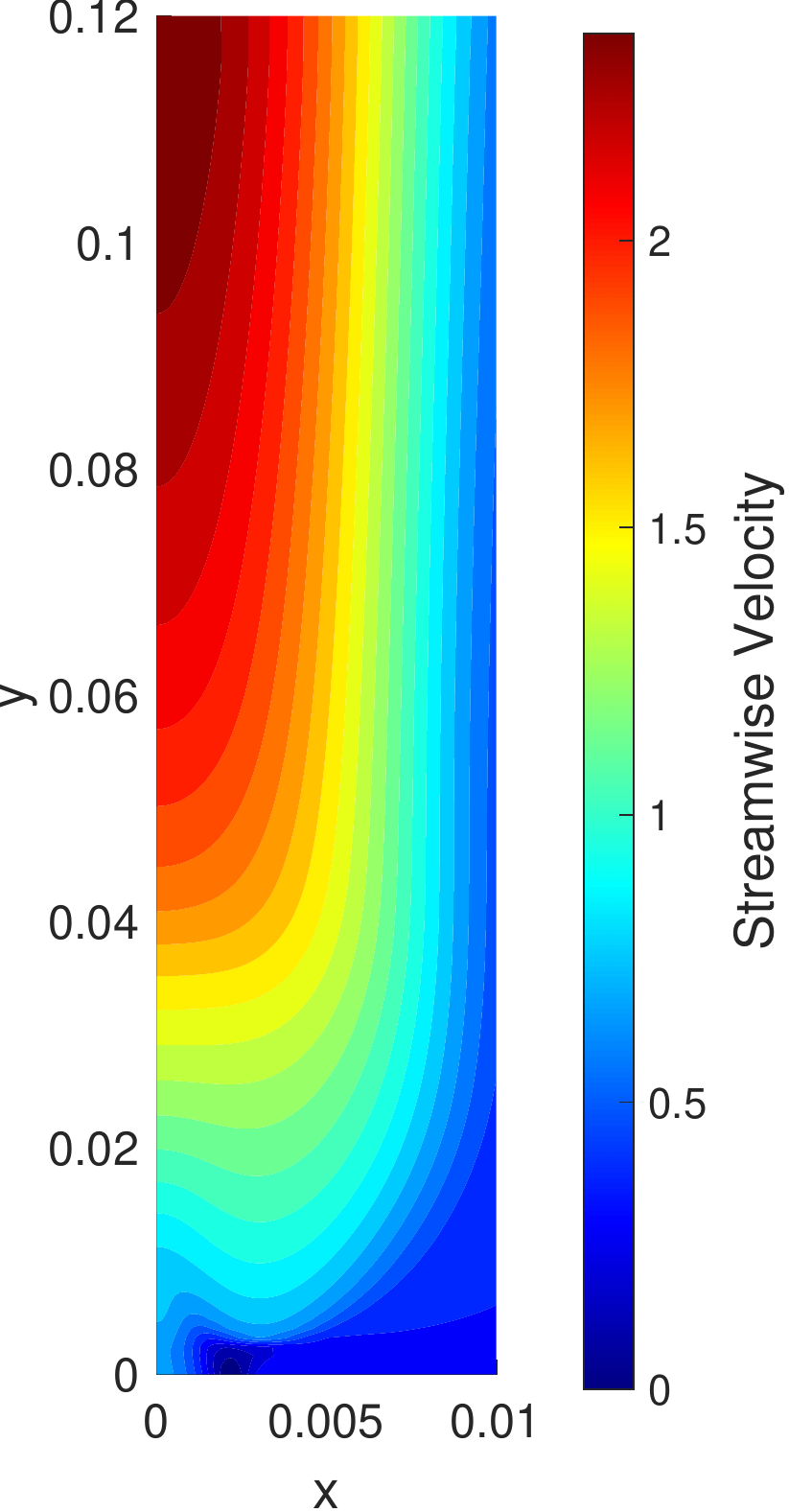}
	\caption{Representative snapshot related to the analysed database. Variables plotted from left to right: mixture fraction, temperature and streamwise velocity . \label{fig:snaps}}
\end{figure}

The dataset representative for the numerical simulation considered for HODMD is a fourth-order tensor formed by 83 variables (the temperature and 82 different species), two spatial dimensions composed by a $150 \times 150$ grid points (along the streamwise and normal direction, respectively); and $117$ snapshots equidistant in time with $\Delta t = 10^{-3}$. Hence, in tensor form, the dimension of the original database is $83\times 150 \times 150 \times 117$. The order of magnitude of the variables ranges from $10^{2}$ (temperature) to $10^{-11}$  (RALD3G, the smallest radical computed). This means that there are $13$ orders of magnitude difference, highlighting the importance of scaling the variables. The variables have been scaled using two different scaling methods, auto scaling and range scaling. In Fig. \ref{fig:snaps}, a representative snapshot is plotted in order to visualise the aforementioned database.

For the PCA-based feature selection, the training matrix $\mathbf{X}$ consists of 22 snapshots, also taking into account the $83$-dimensional thermochemical space.
As in the HODMD analysis, prior to the application of PCA for variable selection, the columns of $\mathbf{X}$ were centered with their mean and scaled with their standard deviation (auto scaling), respectively, because of the different units and ranges of the variables being taken into account. 
Subsequently, the B2 backward elimination algorithm was applied considering a final number of selected variables ranging in the interval $[2, 81]$ (i.e., it was applied imposing each time a different number of variables to retain) and, in particular, first projecting on the standard PCA basis $\mathbf{A}_{q}$, eq. (\ref{eq:projection}),(i.e., applying the standard B2 algorithm), and then rotating the eigenvectors via Varimax (i.e., considering $\mathbf{A}_{q}^{r}$, eq. (\ref{eq:rotatedPC}), and the B2r selection algorithm).

\section{HODMD modelling reactive flows\label{sec:Results}}

The application of HODMD algorithm to develop a reduced order model (ROM) involves a three-step algorithm: the data are pre-processed using different scaling methods at step 1, then, a dimensionality reduction is carried out using HOSVD at step 2, finally the main dynamics associated to the patterns describing the flow are identified at step 3 using HODMD, and a ROM is developed using this information. Two different cases (obtained in step 2) are compared for each type of preprocessing technique used in step 1: (i) using all the components (retaining all the SVD component modes) and (ii) reducing the number of components as explained in eq. (\ref{e26}) in step 2 (only retaining some SVD component modes according to the tunable tolerance $\varepsilon$). It is important to mention that, similarly to PCA, each one of the SVD component modes is related to the number of variables in the database. Therefore, an SVD component mode is composed by the linear combination of the variables studied. The differences found among the various SVD component modes lies in the values of the coefficients forming this linear combination of variables. In what follows, it is important not to confuse the number of components (linear combination of variables), which can be reduced using HOSVD (step 2), with the number of variables, which can be reduced using PCA (as explained in Section \ref{sec:PCAmethod}).

\subsection{HOSVD and truncated variables \label{sec:hosvd}}
 
\begin{figure}
	\centering
	\includegraphics[width = 10 cm]{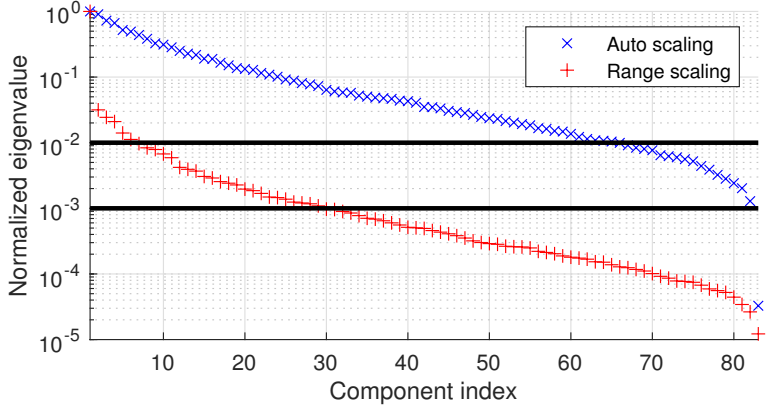}
	\caption{Singular values corresponding to the SVD component modes calculated for the two different scaling methods analysed. The tolerances $\varepsilon$ set are highlighted as horizontal black lines. \label{fig:eigenvalues}}
\end{figure}

The dimensionality reduction carried out using HOSVD to develop a ROM is studied for two different values of the tolerance $\varepsilon$ applied as in eq. (\ref{e26}). As explained before, this tolerance controls the reduction in the spatial and temporal dimensions of the database (SVD spatial and temporal modes), and will also reduce the number of components analysed (SVD component modes) based on their relevant contribution to the dynamical system.  Two tolerances are selected, to compare coarse with fine results, these are: $\varepsilon = 10^{-2}$ and $10^{-3}$.  
Figure \ref{fig:eigenvalues}  compares the number of component retained using the different tolerances using auto scaling and range scaling methods. The value of these tolerances is selected based on the magnitude of the singular values (SVD components). It can be observed that using larger tolerances, the number of components retained when using range scaling is too small (smaller than $5$) to provide accurate results. The smaller tolerance preserves all variables but one (connected to noise) when using auto scaling. This indicates that lowering this limit will not bring useful information.

Table \ref{tab:RRMSE1} shows the RRMSE eq. (\ref{error}) calculated in the reconstruction of the database as in eq. (\ref{e25}): as function of the number of components (SVD component modes) and SVD spatial and temporal modes retained. The HOSVD algorithm has been applied to analyse the dataset with the variables scaled using the range and auto scaling methods, respectively. 
The compression factor, $CF$, as in eq. (\ref{e41}) for the previous dimensionality reduction is presented in Tab. \ref{tab:CF}.
\begin{table}[H]
	\centering
	\begin{tabular}{|c||c|c|c|c|c|c||c|c|c|c|c|c|}
		\hline
		 &\multicolumn{6}{|c||}{Range}&\multicolumn{6}{|c|}{Auto} \\\hline
\hline
		 & $RRMSE_{All}$	&	$RRMSE_{Red}$	& $P_1$ & $P_2$ & $P_3$ & $N$& $RRMSE_{All}$		&	$RRMSE_{Red}$& $P_1$ & $P_2$ & $P_3$& $N$\\
		\hline
		$\varepsilon = 10^{-2}$ & $2.51$\% & $2.82$\% &	$6$ & $7$ & $5$ & $5$ & $2.03$\% & $2.34$\% & $65$ & $40$ & $32$ &$42$\\
		\hline
		$\varepsilon = 10^{-3}$& $0.34$\%	& $0.43$\% & $29$ & $26$ & $19$ & $18$ &$0.18$\% & $0.18$\% & $82$ & $59$ & $48$ & $67$\\
		\hline
	\end{tabular}
	\caption{RRMSE eq. (\ref{error}) (in percentage) comparing the original database to the reconstruction of the reduced tensor, retaining $P_2$, $P_3$, and $N$ spatial (in $x$ and $y$) and temporal SVD modes, respectively. The case {\it All} considers all the components ($P_1=83$), while the case {\it Red} considers the $P_1$ components (SVD component modes) presented in the table. The database is pre-processed as in eq. (\ref{e24}) using the range and auto scaling methods (denoted as Range and Auto, respectively).  
\label{tab:RRMSE1}}
\end{table}
\begin{table}[H]
	\centering
	\begin{tabular}{|c||c|c||c|c|}
		\hline
		&\multicolumn{2}{|c||}{Range}&\multicolumn{2}{|c|}{Auto} \\\hline\hline
		 & All & Red & All & Red \\\hline
		$\varepsilon = 10^{-2}$	& 	$9.18 \times 10^{3}$	& $5.55 \times 10^{4}$	&	$48.72$	& $62.15$	\\
		\hline
		$\varepsilon = 10^{-3}$	& 	$2.9 \times 10^2$	&	$8.11 \times 10^2$	&	$13.84$	&	$14.01$	\\
		\hline
	\end{tabular}
	\caption{Compression factor ($CF$) eq. (\ref{e41}) of the original tensor after applying HOSVD in the databases presented in Tab. \ref{tab:RRMSE1}.  \label{tab:CF}}
\end{table}
 
The RRMSE increases with the tolerance level, as expected. With higher tolerances, the reduction carried out by the HOSVD algorithm is larger, resulting in a larger $CF$. As expected, the RRMSE is larger when the number of components are reduced (Red), as well as the $CF$ increases in this case, especially using the case with the coarse tolerance and the range scaling method. In this case, the $CF$ increases by one order of magnitude while the RRMSE is still maintained below $3$\% (increasing only $0.3$\% compared to the case with all the components (All)). 
 
Comparing the two variable scaling methodologies, the reconstruction of the tensor is slightly better performed when using auto scaling method compared to range scaling, because the former method provides a uniform weight to the variables. However, the compression factor is also much larger when the variables are scaled with range scaling method than with auto scaling. In other words, the number of SVD modes retained for a similar tolerance is larger with auto scaling, thus the compression factor is smaller. 
Similar results were identified when performing a PCA analysis, as described in Ref. \cite{PCA}. 

The main goals of the proposed model is to provide the best reconstruction of the original dataset using the smallest amount of information possible. Based on this idea, the range scaling appears particularly promising because the order of magnitude or the RRMSE is one or two orders larger in the former case, while the rise in the RRMSE error is only reflected in the first decimal digit. More specifically, comparing the results obtained in both cases,  the RRMSE when using range scaling is $\sim 0.3-0.4$\% larger than when using auto scaling in all the cases (the RRMSE is always maintained smaller than $3$\% and $1$\% for the coarse and fine tolerances, respectively), while the $CF$ is $\sim 20-892$\% larger when using range than auto scaling, as presented in Tab. \ref{tab:Comparison}. This result suggests that reducing the number of components and using the range scaling method, it is possible to obtain an accurate ROM (with RRMSE smaller than $3$\% and $1$\% for the coarse and fine tolerances, respectively) with a high dimensionality reduction.
\begin{table}[H]
	\centering
	\begin{tabular}{|c||c|c|}
		\hline
		&\multicolumn{2}{|c|}{$DIFF$} \\\hline\hline
		 & All & Red  \\\hline
		$\varepsilon = 10^{-2}$	& 	$187$\%	& $892$\%		\\
		\hline
		$\varepsilon = 10^{-3}$	& 	$20$\%	&	$57$\%		\\
		\hline
	\end{tabular}
	\caption{Comparison of the $CF$ obtained in the range and auto scaling methods as presented in  Tab. \ref{tab:CF}. The rise in magnitude is calculated as $DIFF=(XX_{range}-XX_{auto})/XX_{auto} \times 100$, with $XX_{range}$ and $XX_{auto}$ as the $CF$ calculated in the range or auto scaling methods, respectively.\label{tab:Comparison}}
\end{table}

Finally, to compare the benefit of each pre-processing method in the reconstruction of the flow (quantified with the RRMSE), when using range and auto scaling methods, it is necessary to adjust the tolerance to have the same compression factor in both cases. Hence,  we select $\varepsilon = 10^{-3}$ and $\varepsilon= 7\times 10^{-2}$ for range and auto scaling, respectively, to build a reduced-dimensionality database, which is considered as a ROM.   The selection of these tolerances gives an RRMSE of $13.11\%$ and $0.34\%$ for auto and range scaling methods respectively, and the corresponding associated compression factor is $377$ and $290$ for the case considering all the components ($All$). The reason of this choice is based on maintaining the largest number of components for the range scaling method (fine tolerance), increasing the accuracy of the ROM. Hence, we adapt the number of components selected in the auto scaling method, as function of the previous choice. This reduced database will be used (as part of the HODMD algorithm) for the spectral analysis presented in the following section.

\subsection{HODMD for the spectral analysis \label{sec:efHODMD}}

\begin{figure}
	\centering
	\includegraphics[width = 9 cm]{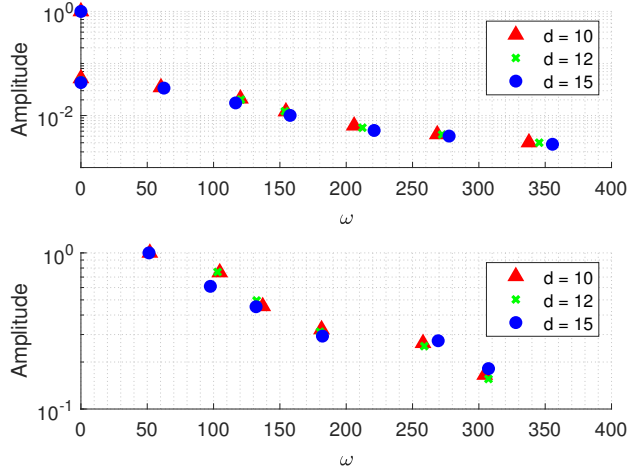}
	\caption{Frequencies vs. amplitudes corresponding to the DMD modes calculated in the combustion database. \textbf{Top:} Range scaling. \textbf{Bottom:} Auto scaling. \label{fig:spectrum}}
\end{figure}

HODMD is applied to identify the main patterns and associated frequencies driving the flow dynamics. To calibrate the algorithm, several values of $d$ have been used, with the aim at ensuring the robustness of the method, these are $d = 10$, $12$ and $15$, which are values close to $11$, a value proportional to the number of snapshots that is calculated  following the calibration tips presented in Ref. \cite{HODMDbook} as $K/10=117/10\simeq11$. The tolerances used for developing the ROM have been selected in the previous section (to maintain a similar compression factor in both the range and auto scaling methods):  $\varepsilon = 10^{-3}$ for range scaling and $\varepsilon= 7\times 10^{-2}$ for auto scaling. Figure \ref{fig:spectrum} shows the frequencies as a function of the amplitudes of the DMD modes identified with the different tolerances. The top figure shows the analysis for range scaling, and the bottom figure, for auto scaling. The spectrum calculated is similar in the analysis carried out in the two databases. 
The spectrum shows a clear periodic solution with a principal mode, the one with the biggest amplitude and frequency $\omega_p$, and the harmonics, modes with lower amplitude and frequency $\omega_n = n \times \omega_p$, being $n \in \mathbb{N} >1$. The principal mode using range scaling is $\omega_p^{RANGE} \simeq 60$, and the one using auto scaling is $\omega_p^{AUTO} \simeq 50$. The reason for this difference lies in the different variables selected as dominant in each one of the scaling methods, which is explained in detail below. A similar spectrum (only one is shown for the sake of brevity) has been obtained for the case retaining all the components (All) and retaining a reduced number of components (Red), as previously presented in Tab. \ref{tab:RRMSE1}. This is because, as explained in section \ref{sec:HODMDcombustion}, the DMD-d algorithm is applied to the temporal matrix, which is not affected by the reduction of the components.

The RRMSE error eq. (\ref{error}) is calculated for the reconstruction of the tensor using the DMD expansion (\ref{ab00}) with the DMD modes presented in Fig. \ref{fig:spectrum} (plus their complex conjugate), representing a purely periodic solution. The parameter $d=12$ has been set as the reference to illustrate this behaviour (although similar results are obtained for $d=10$ and $15$). Table \ref{tab:RRMSEhodmd} presents this error for the cases  All and Red (retaining all and a reduced number of components, respectively). The differences found in this error when HODMD is applied over the database containing all the components and the one with a reduced number of components is small, in good agreement with the results obtained in the previous section. This suggests that the influence of reducing the number of components on the flow dynamics is small. On the other hand, the error is much smaller when using range scaling (error smaller than $1$\%) than auto scaling (error $\sim 24$\%), suggesting that the periodic behaviour identified by HODMD is driven by the main variables (identified in range scaling). This fact indicates that using HODMD to identify the flow dynamics in a database formed by a large number of variables with similar weight (as when using auto scaling) could mask the value of the leading flow frequency, or could alter the proper reconstruction of the original flow (due to the presence of some noise or some error assumed in the frequency calculations).   This is in good agreement with the two different values of leading frequencies identified in the analyses carried out in the database scaled with range and auto scaling methods ($\omega_p^{AUTO} \simeq 50$ and $\omega_p^{RANGE} \simeq 60$). 
\begin{table}
	\centering
	\begin{tabular}{|c|c|c|}
		\hline
		& Range	&	Auto \\
		\hline
		$RRMSE_{All}$	& 	$0.71$\%	&	$24.71$\%	\\
		\hline
		$RRMSE_{Red}$ & 	$0.75$\%	&	$26.06$\%	\\
		\hline
	\end{tabular}
	
	\caption{RRMSE of the reconstructed tensor (in percentage) for $d=12$. Two different scaling methods have been used. \label{tab:RRMSEhodmd}}
\end{table}

\begin{figure}
	\centering
	\includegraphics[width = 10 cm]{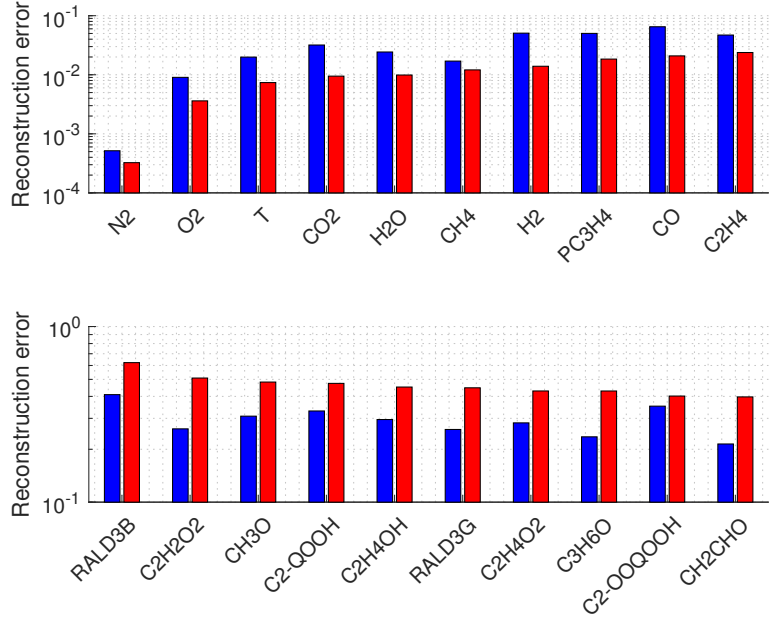}
	\caption{Reconstruction error for the 10 best reconstructed variables with range scaling (top) and the 10 worst reconstructed variables with range scaling (bottom). Blue: Auto scaling. Red: Range scaling. \label{fig:RRMSEbar}}
\end{figure}

Finally, Fig. \ref{fig:RRMSEbar} shows the reconstruction error for the 10 variables with minimum (top) and maximum (bottom) RRMSE identified with range scaling, and the results are compared with the RRMSE obtained for the same variables using auto scaling. It can be observed how range scaling reconstructs with smallest error the main variables, while the largest error is obtained in the reconstruction of the radicals. On the contrary, since auto scaling  provides a uniform reconstruction (see details in Ref. \cite{PCA}), the error for the main variables is larger than when using range scaling, while the inverse holds for radicals.

\subsection{Robustness of the modelling tools \label{sec:robustness}}

\begin{figure}
	\centering
	\includegraphics[width = 0.48\linewidth]{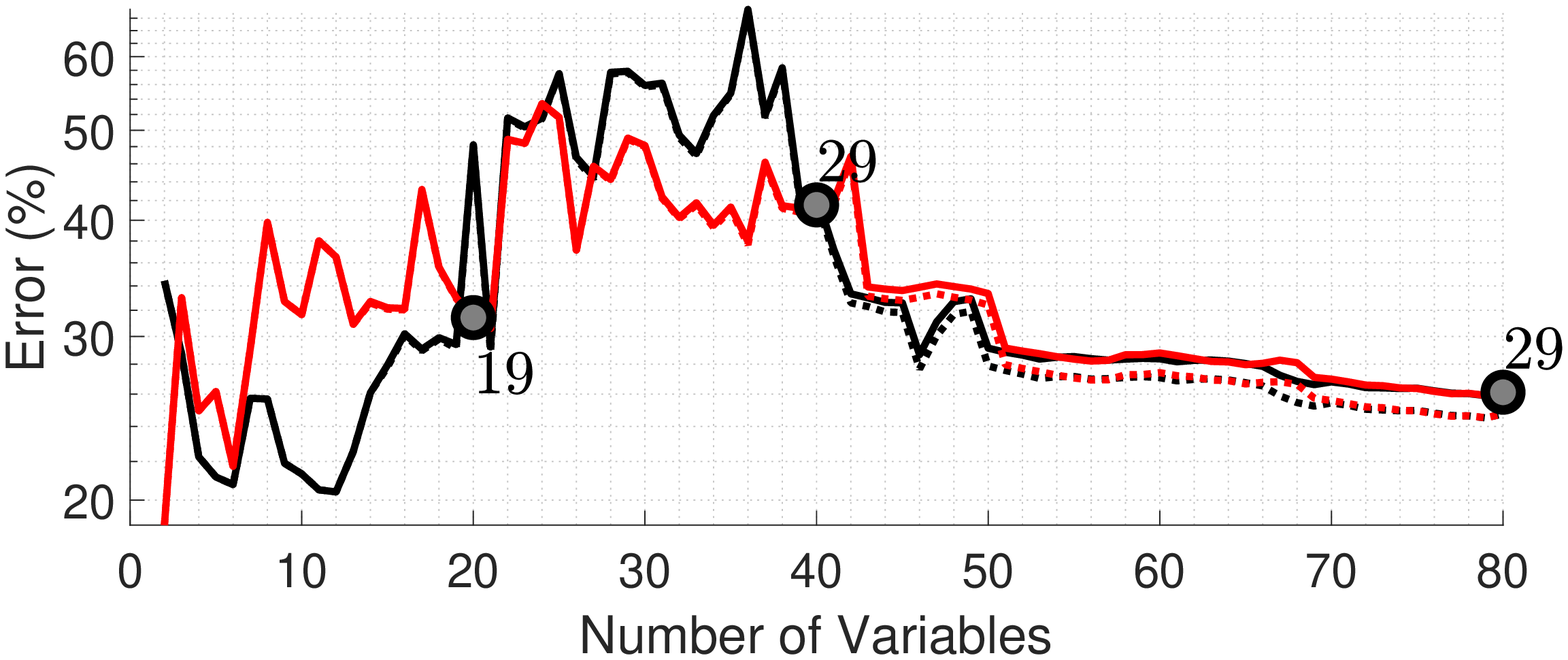}\label{fig:RA_a}
	\includegraphics[width = 0.48\linewidth]{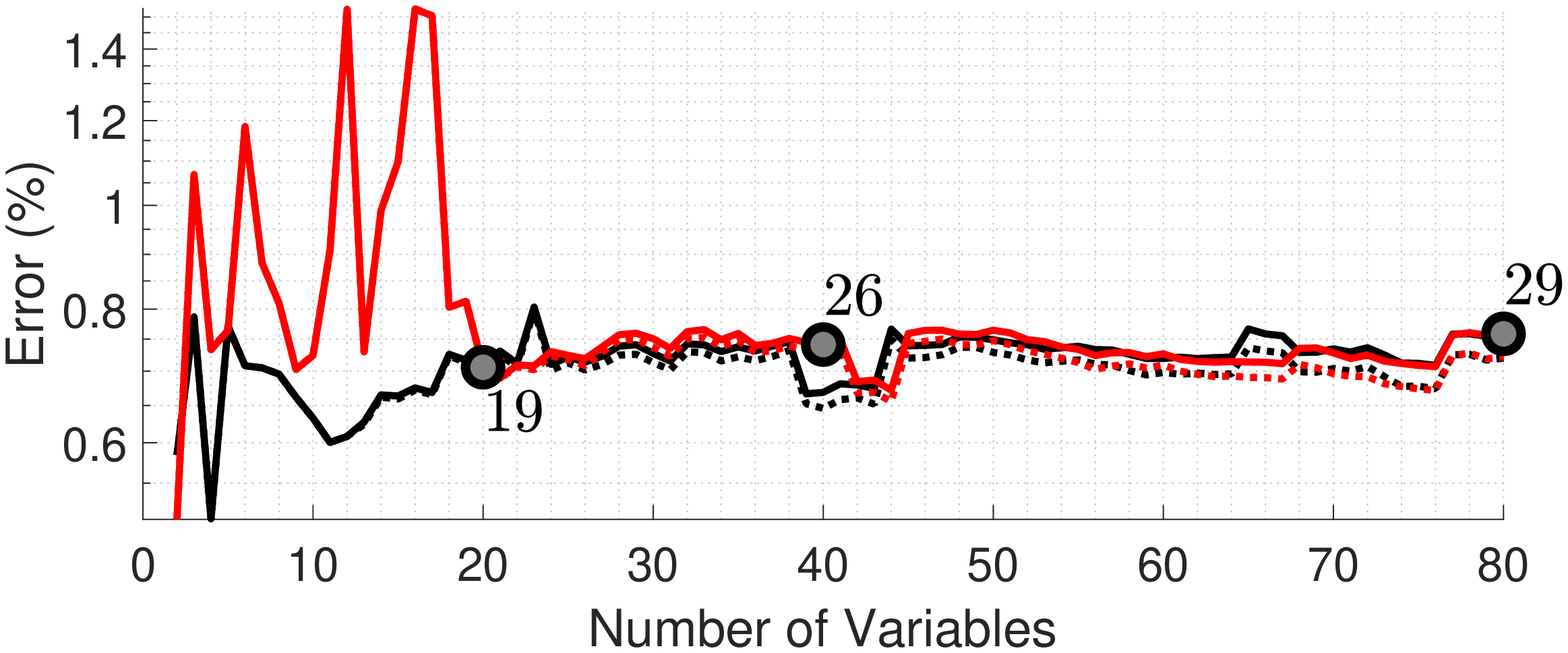}\label{fig:RA_b}
	\caption{Reconstruction error for different variable selection. \textbf{Black:} B2 method variable selection. \textbf{Red:} B2 with rotation method variable selection. \textbf{Dashed line:} All components used. \textbf{Solid line:} Reduced components. \textbf{Left:} Auto scaling. \textbf{Right:} Range scaling. The number marked in the grey points show the number of components selected in HODMD for the specific number of variables presented in the x-label. \label{fig:RRMSE_RA}}
\end{figure}

This section shows the robustness of the ROM previously developed, comparing the RRMSE in the reconstruction as function of number of variables analysed and the number of components. To select the number of variables, PCA is applied over the original database using two methods, B2 and B2 with rotation. The number of variables are selected progressively one by one, resulting in new databases. HODMD is then applied progressively over each one of these databases, calculating the RRMSE for the reconstruction of the database. Moreover, range and auto scaling methods are used in the first step of the HODMD analysis for each one of the new databases. Figure \ref{fig:RRMSE_RA} shows the variation of the RRMSE as function of the number of variables (from PCA) and the number of components (from HODMD) retained. The analysis is carried out in the two databases presented before, the database with all the components (All), and the reduced database (Red). The variable selection method in PCA does not affect significantly the result, although for a small number of selected variables, B2 method decreases the RRMSE. However, the type of scaling method strongly affects the results. The RRMSE for the reconstruction of the flow fluctuates as function of the number of variables in both cases. The error varies from $20-65$\% when using auto scaling and from $0.6-1.5$\% for the range scaling. Additionally, the reconstruction error stabilises using a smaller number of variables with range scaling, suggesting once more the suitability of this scaling strategy to capture the main flow features. Regarding the number of components, which are marked in the points highlighted in grey in the figure, the results show that when increasing the number of variables, reducing the number of components does not increases the RRMSE to a large extent. Hence, when the number of variables is increased, reducing the components may be a good strategy to reduce the data dimensionality of the model while maintaining  a good reconstruction.

\begin{figure}
	\centering
	\includegraphics[width = 0.48\linewidth]{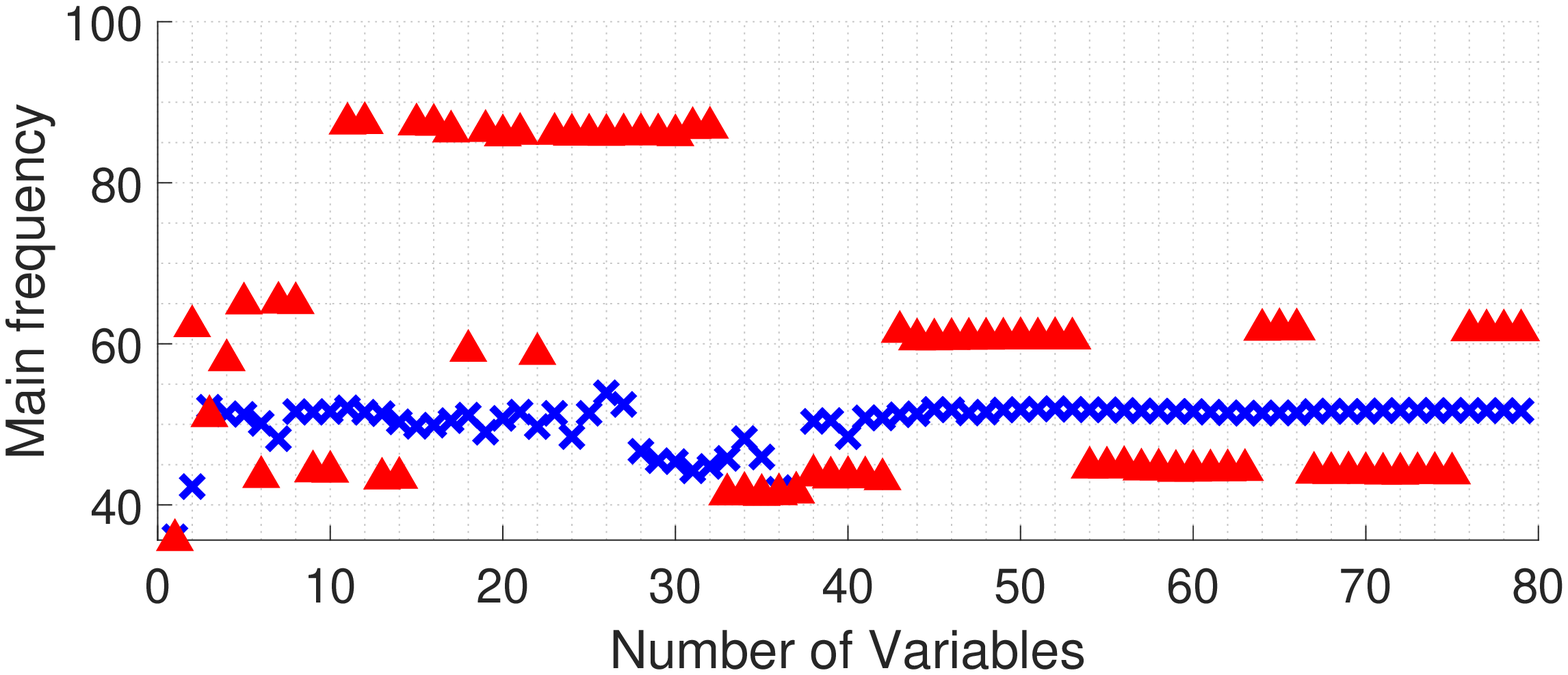}
	\includegraphics[width = 0.48\linewidth]{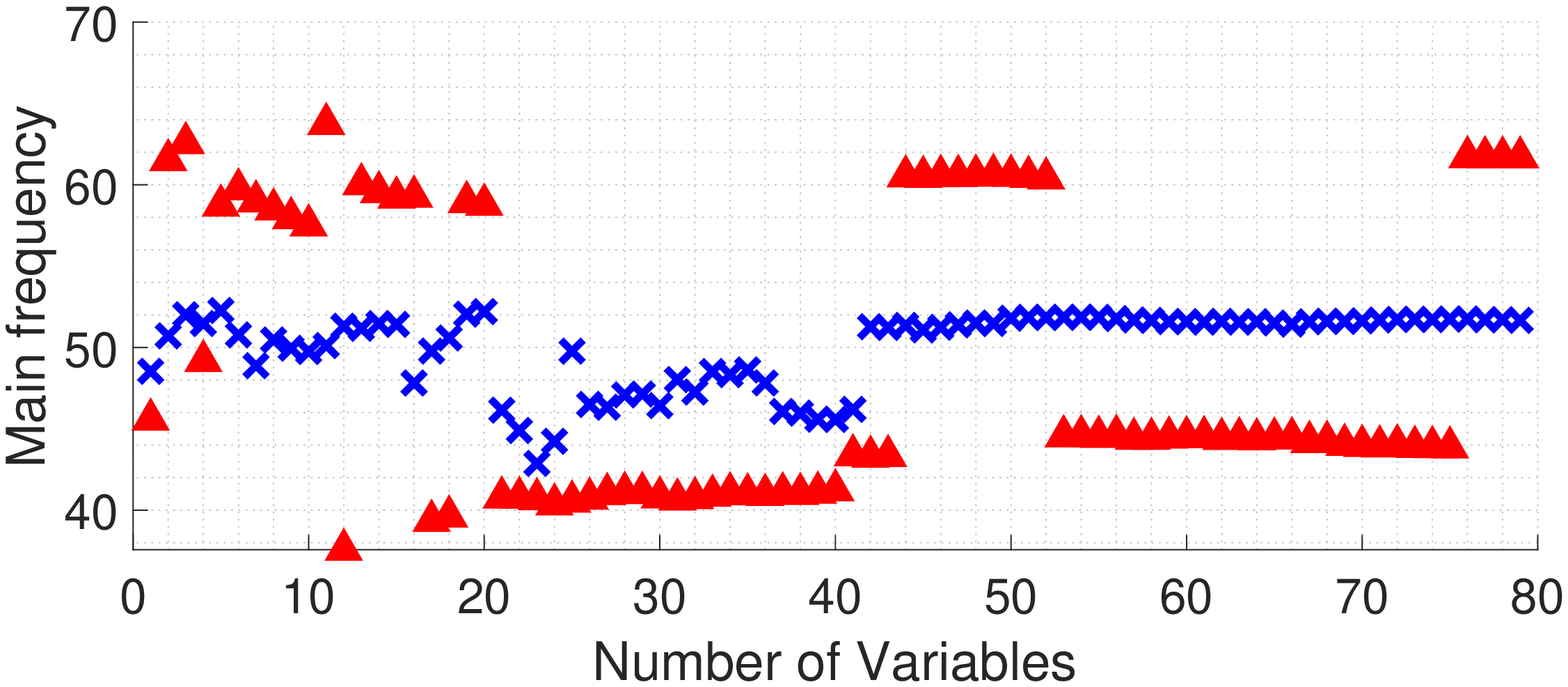}
	\caption{Main frequency vs. number of variables analysed via HODMD. \textbf{Left:}  B2 Variable selection.\textbf{Right:}  B2 with rotation Variable selection. \textbf{Blue crosses:} auto scaling. \textbf{Red triangles:} range scaling. \label{fig:Dynamic}}
\end{figure}

Finally, the main frequency calculated by HODMD as function of the number of variables is plotted in Fig. \ref{fig:Dynamic}. Since the data analysed represent the transient regime of a numerical simulation (which also increases the difficulty of the analysis presented in this work), eq. (\ref{e42}) is used to set the leading modes in each case. The variable selection cases B2 and B2 with rotation are studied in the databases scaled using auto scaling and range scaling methods. Depending on the number of variables, the frequency changes between $\sim 40$, $\sim 60$ and $\sim 90$ when the database is scaled using range method in both selection methods.  Nevertheless, this frequency value is stable for auto scaling, with $\omega \sim  50$. This suggests once more that there is more than one dynamics driving the flow motion (actually, the method identifies three main dynamics) and the range scaling can prioritize them properly. When there is a small number of variables selected, the value of the main frequency oscillates for range scaling. This phenomenon stabilises when the number of variables is increased, hence the method identifies an averaged value for the frequency of the main variables selected.
These results suggest once more, that the main dynamics of the main variables and radicals are driven by different frequencies, at least in the transient regime of the numerical database analysed. 

\section{Conclusions \label{sec:conclusions}}

This article introduces a new application of the multi-dimensional HODMD algorithm, adapted to develop ROMs in reactive flows. This application is based on a new methodology, fully data-driven, consisting on three main steps. The first step focuses on pre-processing the database analysed, using techniques (generally used in machine learning), which consist on centering and scaling each variable. On the second step, HOSVD is applied to reduce the data dimensionality of the database analysed. With this algorithm, the computational cost (time and memory requirements) of the next step thoroughly decreases. Finally, the third step uses the classical HODMD algorithm.

The effect of reducing the number of components is studied in detail measuring the accuracy of the ROM via the relative RMS error (RRMSE) from the reconstruction of the original database. Two different tolerances have been set (coarse and fine), to study two different cases: maintaining all the components in the model or reducing this number of components via SVD. In both cases, the reconstruction error is slightly smaller when using the auto scaling method, although the compression factor is much larger when using range scaling. The reason of this performance is that auto scaling method, gives all the variables the same importance (i.e., temperature and pressure will be considered as important as the radicals), retaining a bigger number of components for a similar tolerance. On the contrary, range scaling identifies the main variables of the flow, hence providing more accurate flow reconstructions using a smaller number of components. Additionally, reducing the number of components composing the model strongly increases the compression factor, while the RRMSE is slightly increased compared to the case with all the variables. This result suggests that using range scaling method and reducing the number of components can be considered as a good strategy to develop a ROM.

HODMD is applied to the previous reduced-dimensionality databases, finding that the main dynamics driving the flow motion is periodic. The main frequency identified using the range and scaling methods is different. The RRMSE for the reconstruction is smaller than $1$\% in the case scaled with range, while this error is $\sim 24$\% when using the auto scaling method. The variations in the reconstruction error is small for the cases reducing and maintaining all the components. The differences found between these errors suggests that using the range scaling, the dynamics driving the main variables are well captured by HODMD.

The robustness of the method to identify the main flow dynamics, has been tested by applying HODMD to different number of variables.  These variables have been selected using PCA with two variable selection methods, B2 and B2 with rotation. The RRMSE of the reconstructed field fluctuates as function of the number of variables as $\sim 0.6-1.5$ and $\sim 20-65$ for the range and auto scaling methods, respectively. Moreover,  this reconstruction error stabilises when using a smaller number of variables in the range scaling method. It is also found that reducing the number of components, does not increase the RRMSE of the reconstruction to a large extent when the error is stabilized. These results suggest once more that using the range scaling method and reducing the number of components is the best option to identify the main flow dynamics and to develop a ROM, with a high dimensionality reduction (compression factor).

Finally, the coupling of the HODMD and the variable selection using PCA suggests that a set of variables are adequate to model the main dynamics of the whole system. This has an impact on Feature selection and on the reduction of the computational cost associated with very massive data.

\section*{Acknowledgments}
AC and SLC acknowledge the grant PID2020-114173RB-I00 funded by MCIN/AEI/10.13039/501100011033.  AC also acknowledges the support of Universidad Politécnica de Madrid, under the programme ‘Programa Propio’. GDA acknowledges the support of the Fonds National de la Recherche Scientifique (FRS-FNRS) through a FRIA fellowship. AP acknowledges funding from the European Research Council (ERC) under the European Union’s Horizon 2020 research and innovation programme under grant agreement No 714605. 

\bibliographystyle{unsrt}
\bibliography{sample.bib}

\begin{thebibliography}{10}

\bibitem{cant2002high}
Stewart Cant.
\newblock High-performance computing in computational fluid dynamics: progress
  and challenges.
\newblock {\em Philosophical Transactions of the Royal Society of London.
  Series A: Mathematical, Physical and Engineering Sciences},
  360(1795):1211--1225, 2002.

\bibitem{parente2011investigation}
Alessandro Parente, JC~Sutherland, Bassam~B Dally, Leonardo Tognotti, and
  PJ~Smith.
\newblock Investigation of the mild combustion regime via principal component
  analysis.
\newblock {\em Proceedings of the Combustion Institute}, 33(2):3333--3341,
  2011.

\bibitem{bellemans2018feature}
Aur{\'e}lie Bellemans, Gianmarco Aversano, Axel Coussement, and Alessandro
  Parente.
\newblock Feature extraction and reduced-order modelling of nitrogen plasma
  models using principal component analysis.
\newblock {\em Computers \& chemical engineering}, 115:504--514, 2018.

\bibitem{Coussement2013}
Axel Coussement, Olivier Gicquel, and Alessandro Parente.
\newblock Mg-local-pca method for reduced order combustion modeling.
\newblock {\em Proceedings of the Combustion Institute}, 34(1):1117--1123,
  2013.

\bibitem{PCA}
Alessandro Parente and James~C Sutherland.
\newblock Principal component analysis of turbulent combustion data: Data
  pre-processing and manifold sensitivity.
\newblock {\em Combustion and flame}, 160(2):340--350, 2013.

\bibitem{isaac2014reduced}
Benjamin~J Isaac, Axel Coussement, Olivier Gicquel, Philip~J Smith, and
  Alessandro Parente.
\newblock Reduced-order pca models for chemical reacting flows.
\newblock {\em Combustion and flame}, 161(11):2785--2800, 2014.

\bibitem{parente2009identification}
Alessandro Parente, James~C Sutherland, Leonardo Tognotti, and Philip~J Smith.
\newblock Identification of low-dimensional manifolds in turbulent flames.
\newblock {\em Proceedings of the Combustion Institute}, 32(1):1579--1586,
  2009.

\bibitem{bishop2006pattern}
Christopher~M Bishop.
\newblock Pattern recognition.
\newblock {\em Machine learning}, 128(9), 2006.

\bibitem{jolliffe2016principal}
Ian~T Jolliffe and Jorge Cadima.
\newblock Principal component analysis: a review and recent developments.
\newblock {\em Philosophical Transactions of the Royal Society A: Mathematical,
  Physical and Engineering Sciences}, 374(2065):20150202, 2016.

\bibitem{d2021feature}
Giuseppe D’Alessio, Alberto Cuoci, and Alessandro Parente.
\newblock Feature extraction and artificial neural networks for the on-the-fly
  classification of high-dimensional thermochemical spaces in
  adaptive-chemistry simulations.
\newblock {\em Data-Centric Engineering}, 2, 2021.

\bibitem{d2020adaptive}
Giuseppe D’Alessio, Alessandro Parente, Alessandro Stagni, and Alberto Cuoci.
\newblock Adaptive chemistry via pre-partitioning of composition space and
  mechanism reduction.
\newblock {\em Combustion and Flame}, 211:68--82, 2020.

\bibitem{d2020impact}
Giuseppe D’Alessio, Alberto Cuoci, Gianmarco Aversano, Mauro Bracconi,
  Alessandro Stagni, and Alessandro Parente.
\newblock Impact of the partitioning method on multidimensional
  adaptive-chemistry simulations.
\newblock {\em Energies}, 13(10):2567, 2020.

\bibitem{d2020analysis}
Giuseppe D’Alessio, Antonio Attili, Alberto Cuoci, Heinz Pitsch, and
  Alessandro Parente.
\newblock Analysis of turbulent reacting jets via principal component analysis.
\newblock In {\em Data Analysis for Direct Numerical Simulations of Turbulent
  Combustion}, pages 233--251. Springer, 2020.

\bibitem{jolliffe1972discarding}
Ian~T Jolliffe.
\newblock Discarding variables in a principal component analysis. i: Artificial
  data.
\newblock {\em Journal of the Royal Statistical Society: Series C (Applied
  Statistics)}, 21(2):160--173, 1972.

\bibitem{jolliffe1973discarding}
Ian~T Jolliffe.
\newblock Discarding variables in a principal component analysis. ii: Real
  data.
\newblock {\em Journal of the Royal Statistical Society: Series C (Applied
  Statistics)}, 22(1):21--31, 1973.

\bibitem{krzanowski1987selection}
Wojtek~J Krzanowski.
\newblock Selection of variables to preserve multivariate data structure, using
  principal components.
\newblock {\em Journal of the Royal Statistical Society: Series C (Applied
  Statistics)}, 36(1):22--33, 1987.

\bibitem{jolliffe2002choosing}
Ian~T Jolliffe.
\newblock Choosing a subset of principal components or variables.
\newblock {\em Principal component analysis}, pages 111--149, 2002.

\bibitem{Schmid10}
Peter~J Schmid.
\newblock Dynamic mode decomposition of numerical and experimental data.
\newblock {\em Journal of fluid mechanics}, 656:5--28, 2010.

\bibitem{LeClaincheSOCO19}
Soledad Le~Clainche.
\newblock An introduction to some methods for soft computing in fluid dynamics.
\newblock {\em Advances in Intelligent Systems and Computing}, 950:557--566,
  2019.

\bibitem{Richecoeur2012}
Franck Richecoeur, Layal Hakim, Antoine Renaud, and Laurent Zimmer.
\newblock {DMD algorithms for experimental data processing in combustion}.
\newblock In {\em {Proceeding of the 2012 Summer Program}}, pages 459--468.
  {Center for Turbulence Research, Stanford University}, December 2012.

\bibitem{Souvick2013}
Souvick Chatterjee, Achintya Mukhopadhyay, and Swarnendu Sen.
\newblock Stability study of laminar flame using proper orthogonal decompostion
  and dynamic mode decomposition.
\newblock In {\em n3l-Int’l Summer School and Workshop on Non-Normal and
  Nonlinear Effects In Aero-and Thermoacoustics}, page~13, 2013.

\bibitem{Quinlan2014}
John~M. Quinlan and Ben~T. Zinn.
\newblock {\em Transverse Combustion Instabilities: Modern Experimental
  Techniques and Analysis}.
\newblock 2014.

\bibitem{Huang2016}
Cheng Huang, William~E. Anderson, Matthew~E. Harvazinski, and Venkateswaran
  Sankaran.
\newblock Analysis of self-excited combustion instabilities using decomposition
  techniques.
\newblock {\em AIAA Journal}, 54(9):2791--2807, 2016.

\bibitem{Motheau2014}
Emmanuel Motheau, Franck Nicoud, and Thierry Poinsot.
\newblock {Mixed acoustic-entropy combustion instabilities in gas turbines}.
\newblock {\em {Journal of Fluid Mechanics}}, 749:542-- 576, 2014.

\bibitem{AbouTaouk2015}
Abdallah Abou-Taouk, S.K. Sadasivuni, Daniel Lörstad, Ghenadie Bulat, and
  Lars-Erik Eriksson.
\newblock Cfd analysis and application of dynamic mode decomposition for
  resonant-mode identification and damping in an sgt-100 dle combustion system.
\newblock In {\em Proceedings of the 7th European Combustion Meeting}, 03 2015.

\bibitem{Ghani2015}
Abdulla Ghani, Thierry Poinsot, Gicquel L.Y.M., and Gabriel Staffelbach.
\newblock Les of longitudinal and transverse self-excited combustion
  instabilities in a bluff-body stabilized turbulent premixed flame.
\newblock {\em Combustion and Flame}, 162:4075--4083, 09 2015.

\bibitem{grenga2018dynamic}
Temistocle Grenga, Jonathan~F MacArt, and Michael~E Mueller.
\newblock Dynamic mode decomposition of a direct numerical simulation of a
  turbulent premixed planar jet flame: convergence of the modes.
\newblock {\em Combustion Theory and Modelling}, 22(4):795--811, 2018.

\bibitem{grenga2020dynamic}
T~Grenga and ME~Mueller.
\newblock Dynamic mode decomposition: A tool to extract structures hidden in
  massive datasets.
\newblock In {\em Data Analysis for Direct Numerical Simulations of Turbulent
  Combustion}, pages 157--176. Springer, 2020.

\bibitem{LeClaincheVegaComplexity18}
Soledad Le~Clainche and Jos{\'e}~M Vega.
\newblock Analyzing nonlinear dynamics via data-driven dynamic mode
  decomposition-like methods.
\newblock {\em Complexity}, 2018, 2018.

\bibitem{LeClaincheVega17}
Soledad Le~Clainche and Jos{\'e}~M Vega.
\newblock Higher order dynamic mode decomposition.
\newblock {\em SIAM Journal on Applied Dynamical Systems}, 16(2):882--925,
  2017.

\bibitem{Corrochano}
Adri{\'a}n Corrochano, Donnatella Xavier, Philipp Schlatter, Ricardo Vinuesa,
  and Soledad Le~Clainche.
\newblock Flow structures on a planar food and drug administration (fda) nozzle
  at low and intermediate reynolds number.
\newblock {\em Fluids}, 6(1):4, 2021.

\bibitem{LeClaincheetalAIAA17}
Soledad Le~Clainche Martinez, Francisco Sastre, José~M. Vega, and Velazquez
  Angel.
\newblock {\em Higher order dynamic mode decomposition applied to post-process
  a limited amount of noisy PIV data}.

\bibitem{LeClaincheetalJAircraft18}
Soledad Le~Clainche, Rub{\'e}n Moreno-Ramos, Paul Taylor, and Jos{\'e}~M Vega.
\newblock New robust method to study flight flutter testing.
\newblock {\em Journal of Aircraft}, 56(1):336--343, 2019.

\bibitem{LeClaincheetalJFM2020}
S~Le~Clainche, Daulet Izbassarov, M~Rosti, Luca Brandt, and Outi Tammisola.
\newblock Coherent structures in the turbulent channel flow of an
  elastoviscoplastic fluid.
\newblock {\em Journal of Fluid Mechanics}, 888, 2020.

\bibitem{kaiser1958varimax}
Henry~F Kaiser.
\newblock The varimax criterion for analytic rotation in factor analysis.
\newblock {\em Psychometrika}, 23(3):187--200, 1958.

\bibitem{LeClaincheVegaSoria17}
Soledad Le~Clainche, Jos{\'e}~M Vega, and Julio Soria.
\newblock Higher order dynamic mode decomposition of noisy experimental data:
  The flow structure of a zero-net-mass-flux jet.
\newblock {\em Experimental Thermal and Fluid Science}, 88:336--353, 2017.

\bibitem{HODMDbook}
Jose~Manuel Vega and Soledad Le~Clainche.
\newblock {\em Higher order dynamic mode decomposition and its applications}.
\newblock Academic Press, 2020.

\bibitem{LeClaincheFerrer2018}
Soledad Le~Clainche and Esteban Ferrer.
\newblock A reduced order model to predict transient flows around straight
  bladed vertical axis wind turbines.
\newblock {\em Energies}, 11(3):566, 2018.

\bibitem{LeClaincheVegaPoF17}
Soledad Le~Clainche and Jos{\'e}~M Vega.
\newblock Higher order dynamic mode decomposition to identify and extrapolate
  flow patterns.
\newblock {\em Physics of Fluids}, 29(8):084102, 2017.

\bibitem{HOSVD}
Ledyard~R Tucker.
\newblock Some mathematical notes on three-mode factor analysis.
\newblock {\em Psychometrika}, 31(3):279--311, 1966.

\bibitem{Chenetal12}
K~K Chen, J~H Tu, and C~W Rowley.
\newblock Variants of dynamic mode decomposition: Boundary condition, koopman,
  and fourier analyses.
\newblock {\em {J}. Nonlin. Scien.}, 22(6):887--915, 2012.

\bibitem{Kou}
Jiaqing Kou, Soledad Le~Clainche, and Weiwei Zhang.
\newblock A reduced-order model for compressible flows with buffeting condition
  using higher order dynamic mode decomposition with a mode selection
  criterion.
\newblock {\em Physics of Fluids}, 30(1):016103, 2018.

\bibitem{POLI}
Eliseo Ranzi, Alessio Frassoldati, Roberto Grana, Alberto Cuoci, Tiziano
  Faravelli, Andrew~P Kelley, and Chung~K Law.
\newblock Hierarchical and comparative kinetic modeling of laminar flame speeds
  of hydrocarbon and oxygenated fuels.
\newblock {\em Progress in Energy and Combustion Science}, 38(4):468–501,
  2012.

\bibitem{cuoci2013numerical}
Alberto Cuoci, Alessio Frassoldati, Tiziano Faravelli, and Eliseo Ranzi.
\newblock Numerical modeling of laminar flames with detailed kinetics based on
  the operator-splitting method.
\newblock {\em Energy \& Fuels}, 27(12):7730--7753, 2013.

\end{thebibliography}

\end{document}